\documentclass[letterpaper]{article} 
\usepackage[preprint]{aaai2027}  
\usepackage[hyphens]{url}  
\usepackage{graphicx} 
\usepackage{natbib}  
\usepackage{caption} 
\usepackage{amsmath}
\usepackage{amssymb}
\usepackage{multirow}
\usepackage{booktabs}
\usepackage{algorithm}
\usepackage{algorithmic}

\title{Beyond Varshamov-Tenengolts Codes: Sketch-Guided Neural Decoding for Multiple InDel Errors}
\author{Guangya Hu\textsuperscript{1}, Shu Liu\textsuperscript{2}}
\affiliations{
\textsuperscript{1}Shanghai Jiao Tong University, Shanghai, China\\
\textsuperscript{2}National Key Laboratory of Wireless Communications,\\
University of Electronic Science and Technology of China, Chengdu, China\\
hgyhang@sjtu.edu.cn, shuliu@uestc.edu.cn}
\begin{document}
\maketitle

\begin{abstract}
Correcting multiple insertion and deletion (InDel) errors, which cause severe synchronization drift across data sequences, remains a longstanding challenge in coding theory. Traditional algebraic codes capable of handling multiple InDel errors suffer from stringent code constraints and exponential decoding complexity. We propose an error-correcting framework that imposes concise, higher-order sketch constraints on the codeword and employs a neural Transformer-based decoder for message recovery. To handle synchronization drift, the decoder integrates convolutional layers for local context extraction and gated attention mechanisms for adaptive modulation of attention outputs. On binary InDel channels, we empirically observe that each additional sketch imposed on the codeword extends the decoder's capability to correct one additional error, suppressing the Bit Error Rate (BER) below 1\% under multiple concurrent errors. The decoding time complexity is independent of the number of errors, in contrast to the exponential scaling of traditional algebraic decoders. When evaluated on the more complex Insertion, Deletion, and Substitution (IDS) channels, the framework maintains effective error-correction capabilities. We further adapt this framework to DNA storage channels by incorporating GC-content and homopolymer constraints, validating its practical potential for DNA-based data storage.
\end{abstract}
\section{Introduction}
In modern digital communication and storage systems, ensuring data integrity against noise is a fundamental challenge. While traditional error-correcting codes are highly optimized for localized substitution errors, many emerging applications—such as file synchronization, network packet routing, and particularly DNA-based data storage \cite{ceze2019molecular,sabary2024reconstruction}—are inherently plagued by insertion and deletion (InDel) errors. Unlike substitutions, InDel errors cause severe synchronization drift across the entire sequence, corrupting the positional integrity of the data and making recovery exceedingly difficult.

The foundational problem of constructing codes capable of correcting multiple InDel errors has been a longstanding challenge in coding theory. While the classic Varshamov-Tenengolts (VT) code~\cite{varshamov1965code} elegantly resolves the single-deletion scenario, generalizing it to correct multiple errors has proven highly nontrivial. Although recent theoretical breakthroughs have yielded explicit algebraic constructions for multiple deletions \cite{sima2020optimal,guruswami2021explicit}, their practical deployment remains severely hindered. These algebraic approaches rely on extremely complex code constraints and suffer from a prohibitive lack of efficient decoding algorithms—often exhibiting decoding complexities that scale exponentially with the number of errors $t$, such as $\mathcal{O}(n^{t+1})$ \cite{sima2020optimal}.

To bridge the gap between theoretical error-correction bounds and practical decodability, we propose a deep learning-based error-correcting framework. Rather than explicitly solving intractable algebraic syndromes, we utilize deep neural networks to implicitly learn the inverse mapping from corrupted sequences to original messages, guided by lightweight mathematical constraints. This work fills the vacancy that existing studies merely provide constructions of higher-order constraints and characterize their error-correcting abilities, yet fail to offer explicit decoding algorithms. The main contributions of this paper are summarized as follows:
\begin{itemize}
    \item \textbf{Sketch-Based Encoding with Neural Decoding}: We propose an error-correcting framework that combines concise higher-order Combinatorial Sketches ($f_1, f_2, f_3$) as global structural constraints with a Locality-Aware Hybrid Transformer decoder. The decoder performs direct end-to-end message recovery, bypassing the reconstruction of redundant bits.
    \item \textbf{Synchronization-Aware Architectural Design}: We introduce a Multi-scale Residual Convolutional Embedding (MRCE) and a Query-Dependent Gated Attention (QDGA) mechanism to filter pseudo-alignment noise from sequence shifts. These two modules enhance decoding robustness from both the local and global perspectives and render decoding complexity independent of number of errors.
    \item \textbf{Scalable Multi-Error Correction}: On binary InDel channels, each additional sketch constraint extends the decoder's capability to correct one additional error, suppressing the BER below 1\% under multiple concurrent errors. This scalable trend generalizes to the more complex IDS channel, where augmenting sketches consistently recovers significant performance margins.
    \item \textbf{Extension to DNA Storage with Biological Constraints}: The framework is extended to the quaternary DNA alphabet (A, C, G, T) subject to GC balance and homopolymer constraints, with corresponding adaptations to the input embedding and output loss function, demonstrating its viability for real-world DNA storage systems.
\end{itemize}

\section{Related Works}
The Varshamov--Tenengolts (VT) code was originally introduced for correcting a single asymmetric error~\cite{varshamov1965code} and subsequently became central to single-insertion/deletion correction~\cite{levenshtein1966binary}. Early studies extended this line to synchronization, substitution, and related deletion errors~\cite{calabi1969family,tanaka1976synchronization,sloane2000single,paluncic2012multiple}. For multiple deletions, later explicit constructions have substantially reduced redundancy. For constant $t$, order-optimal $8t\log n+o(\log n)$ redundancy has been achieved for binary $t$-deletion codes~\cite{sima2020optimal}, with encoding and decoding complexities $\mathcal{O}(n^{2t+1})$ and $\mathcal{O}(n^{t+1})$, respectively. For two deletions, an explicit construction matches the best-known existential bound with $4\log n+O(\log\log n)$ redundancy~\cite{guruswami2021explicit}. More recently, low-redundancy constructions have been extended to constant-$q$ alphabets, attaining $5\log n+10\log\log n+3\log q+O(1)$ redundancy~\cite{liu2024qary}. These works focus on worst-case deletion correction through specialized code constructions; scalable decoding for mixed InDel errors remains less explored, motivating our sketch-guided neural approach.

Machine learning has long been explored for communication systems~\cite{ibnkahla2000applications,simeone2018brief,akrout2023domain}. In channel decoding, neural networks have replaced constituent decoders in partitioned polar codes~\cite{cammerer2017scaling}, while RNNs have decoded convolutional and turbo codes~\cite{kim2018communication}. Model-based learned belief propagation improves linear-code decoding~\cite{nachmani2016learning,nachmani2018deep}, whereas model-free designs learn decoding mappings directly~\cite{jiang2019deepturbo}. Transformer-based decoding then introduced syndrome-aware global reasoning through ECCT~\cite{choukroun2022error}; this line has expanded to diffusion-based decoding, quantum error correction, foundation-model pretraining, and joint code learning~\cite{choukroundenoising,choukroun2024deepquantum,choukroun2024foundation,choukroun2024learning}. MM-ECCT used multiple masks to enhance the capability of transformer for error correction codes~\cite{park2025multiple}. CrossMPT further combines cross-attention with message passing~\cite{park2025crossmpt}.  Most of these methods focus on AWGN or substitution-dominated channels, motivating our Transformer-based treatment of synchronization errors.

Recently, deep learning has been increasingly explored to address the unique challenges of IDS channels, particularly in the context of DNA-based data storage. Several works have investigated replacing traditional algebraic components with neural networks. For instance, Guo et al. introduced DoDo-Code~\cite{guo2025dodo}, which leverages deep Levenshtein distance embedding to design high-rate 4-ary single-IDS-correcting codes and perform efficient segment decoding. To facilitate end-to-end training over non-differentiable IDS channels, Guo et al. proposed THEA-Code~\cite{guo2026disturbance}, an autoencoder framework incorporating a differentiable simulated IDS channel and disturbance-based discretization for tailored codeword generation. Furthermore, for established algebraic codes like the VT code, Wei et al. proposed VT-Former~\cite{wei2025vtformer}, which employs a statistic-enhanced Transformer architecture to provide a highly efficient soft-decoding alternative capable of correcting multiple IDS errors. While these methods demonstrate the potential of deep learning for IDS channels, our work uniquely integrates higher-order combinatorial sketches with deep learning to achieve scalable, multi-error correction without relying on the autoencoder generation of unconstrained codebooks.

\section{Preliminaries}
In this section, we briefly introduce the preliminary concepts utilized throughout this paper, including deletion correction codes and the Transformer architecture.

\subsection{Deletion Correction and VT Codes}
A $t$-deletion correcting code is defined as a set of sequences of length $n$ capable of recovering from up to $t$ deletions. We provide a formal condition based on the insertion/deletion (InDel) distance, following~\cite{wu2024constructions}.

\vspace{\baselineskip}
\noindent
\textbf{Definition 1.} Let $\mathbf{a}$ and $\mathbf{b}$ be two sequences. The InDel distance $d_{\mathrm{ID}}(\mathbf{a}, \mathbf{b})$ is the minimum number of insertions and deletions needed to transform $\mathbf{a}$ into $\mathbf{b}$.
\vspace{\baselineskip}

\noindent
\textbf{Definition 2.} A code $\mathcal{C} \subseteq \Sigma^n$ is called a $t$-insertion/deletion-correcting code if every codeword can be uniquely recovered after any combination of at most $t$ insertions and deletions in total.
\vspace{\baselineskip}

\noindent
\textbf{Lemma 1.} A code $\mathcal{C}$ can successfully correct any combination of up to a total of $t$ insertions and deletions if and only if $d_{\mathrm{ID}}(\mathbf{a}, \mathbf{b}) > 2t$ for any two distinct codewords $\mathbf{a}, \mathbf{b} \in \mathcal{C}$.
\vspace{\baselineskip}

For the specific case of correcting a single deletion ($t=1$), the Varshamov-Tenengolts (VT) construction provides an explicit code with asymptotically optimal redundancy. The VT code is formally defined as the set of sequences $\mathbf{c} \in \{0,1\}^n$ that satisfy the checksum equation:
\begin{equation}
    \sum_{i=1}^n i c_i \equiv 0 \pmod{n+1}.
\end{equation}
While the VT code perfectly resolves the single-deletion scenario, generalizing it for $t > 1$ has proven to be a longstanding challenge. Natural attempts to extend the VT construction to handle multiple deletions by employing straightforward higher-order forms, such as $\sum_{i=1}^n i^j c_i \pmod{n^j+1}$ for $j=0,1,2,\cdots$, have not been successful~\cite{brakensiek2017efficient}, failing to guarantee unique decodability even for $t=2$.

\subsection{Transformer}
The Transformer architecture~\cite{vaswani2017attention}, which was initially proposed for machine translation, introduced a novel attention-based approach. It processes the input sequence by incorporating positional embeddings for each element and embedding it into a high-dimensional space.

The embeddings are subsequently propagated through multiple normalized self-attention and feed-forward layers. The self-attention mechanism is based on a trainable associative memory with (key, value) vector pairs, where a query vector $\mathbf{q} \in \mathbb{R}^d$ is matched against a set of $\mathbf{k}$ key vectors using scaled inner products, such that $\text{Attention}(Q, K, V)=\text{Softmax}(d^{-1/2}(QK^T))V$. Here $Q \in \mathbb{R}^{N\times d}, K\in \mathbb{R}^{\kappa \times d}$ and $V\in \mathbb{R}^{\kappa\times d}$ represent the stacked $N$ queries, $\kappa$ keys and values tensors, respectively. Keys, queries, and values are obtained using linear transformations of representations of the sequence's elements, and a multi-head self-attention scheme is deployed by extending the self-attention mechanism to multiple attention heads.

\section{Multi-Deletion Correcting Framework}
In this section, we define the higher-order combinatorial sketches used for sequence encoding and outline the end-to-end workflow of our error-correcting framework.

\subsection{Higher-Order Combinatorial Sketches}
While the original VT construction is optimal for correcting a single deletion, recovering from multiple insertions and deletions necessitates richer structural information. Existing efforts extending VT codes with high-order parity checks often result in complex constructions that lack efficient decoding algorithms \cite{guruswami2021explicit,brakensiek2017efficient,sima2020optimal}.

Instead of relying purely on algebraic decoders, we shift the paradigm by leveraging deep learning. To provide the neural decoder with sufficient structural anchors, we employ up to three higher-order combinatorial sketches, denoted as $f_1(\mathbf{x})$, $f_2(\mathbf{x})$, and $f_3(\mathbf{x})$, for a binary codeword $\mathbf{x} \in \{0, 1\}^n$:
\begin{align}
    f_1(\mathbf{x}) &= \sum_{i=1}^n i x_i \pmod{M_1}, \\
    f_2(\mathbf{x}) &= \sum_{i=1}^n \binom{i}{2} x_i \pmod{M_2}, \\
    f_3(\mathbf{x}) &= \sum_{i=1}^n \binom{i}{3} x_i \pmod{M_3},
\end{align}
where $M_1, M_2$, and $M_3$ are predetermined moduli. Here, $f_1(\mathbf{x})$ represents the first-order position weight, conceptually analogous to the classical VT syndrome, while $f_2(\mathbf{x})$ and $f_3(\mathbf{x})$ introduce second and third-order constraints. 

In traditional coding theory, it is difficult to construct efficient decoding algorithms for these higher-order constraints. In prior literature \cite{guruswami2021explicit}, researchers only presented constructions of higher-order constraints together with their error-correcting capabilities, without providing explicit decoding algorithms. In contrast, our deep neural network serves as an efficient decoder for codes subject to these higher-order sketches constraints.

\subsection{System Pipeline and Decoding Workflow}

\begin{figure}[htbp]
    \centering
    \includegraphics[width=0.9\linewidth]{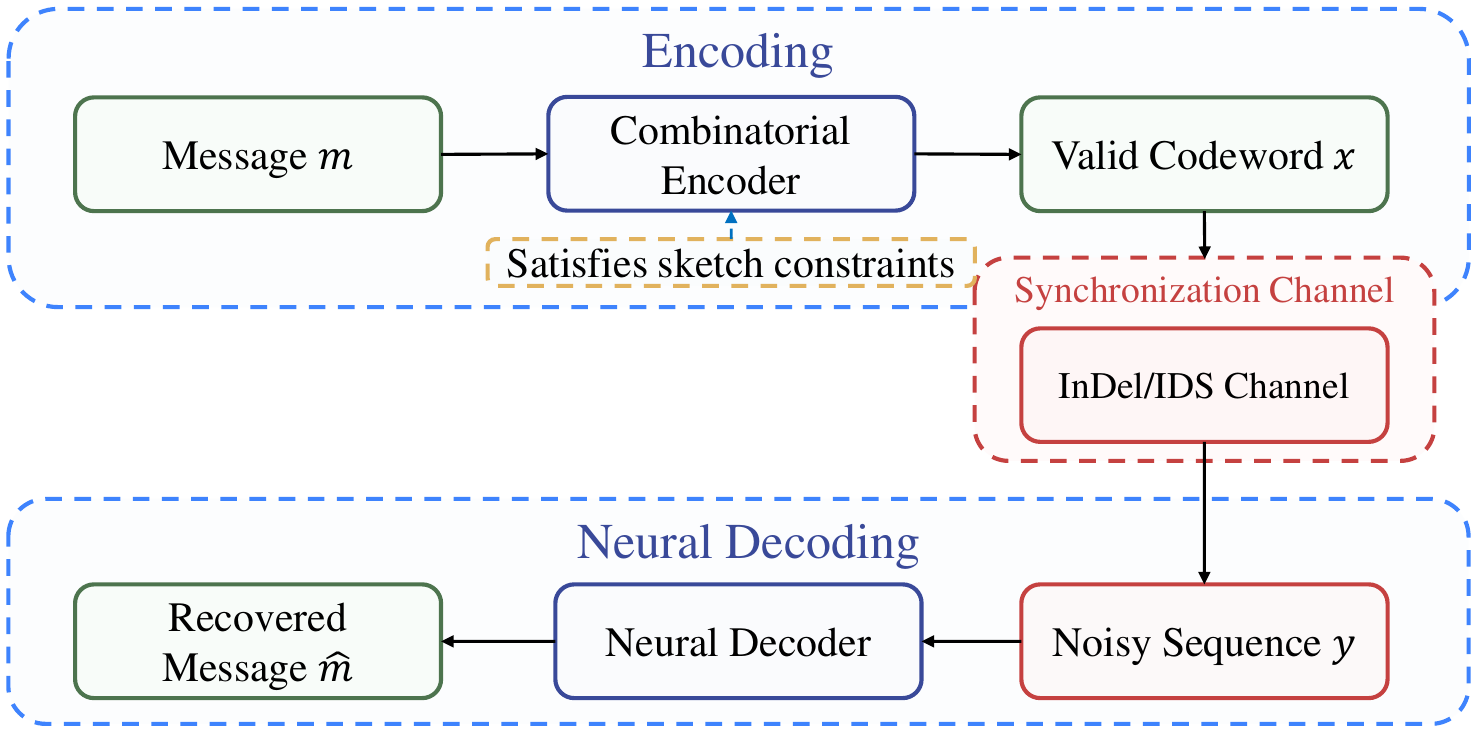}
    \caption{The end-to-end workflow of the proposed system. The encoder searches for an $n$-bit codeword consistent with a $k$-bit message and the selected combinatorial sketches. After transmission through the InDel channel, the neural decoder reconstructs the $k$-bit message from the noisy sequence.}
    \label{fig:system_pipeline}
\end{figure}

The overall architecture of our proposed error-correction system, spanning from message encoding to neural decoding, is illustrated in Figure \ref{fig:system_pipeline}. 

Let $\mathcal{I}\subseteq\{1,\ldots,n\}$ be a fixed set of information positions with $|\mathcal{I}|=k$. Given a $k$-bit message $\mathbf{m}$, the encoder fixes $\mathbf{x}_{\mathcal{I}}=\mathbf{m}$ and searches over assignments to the remaining $n-k$ positions for a codeword $\mathbf{x}$ satisfying the selected sketch constraints. Such an assignment need not exist for every $\mathbf{m}\in\{0,1\}^{k}$; hence, the supported message set may contain fewer than $2^k$ elements and the effective rate may be below the nominal value $k/n$. The codeword-search algorithm, dataset construction procedure, and a Monte Carlo estimate of the empirical rate are provided in the appendix.

Once found, the valid codeword is transmitted through the Insertion/Deletion (InDel) channel. Random insertions and deletions alter its length and shift symbol positions, producing a noisy sequence at the receiver.

The neural decoder receives only this noisy sequence and directly estimates the $k$ information bits. The sketch values are not supplied as separate network inputs; their constraints are embodied in the transmitted codeword. Unlike a full-codeword reconstruction procedure, our decoder does not predict the $n-k$ redundant positions, thereby reducing the output length and focusing the learning objective on message recovery.

\section{Locality-Aware Hybrid Transformer}
In this section, we detail the architecture of our proposed Locality-Aware Hybrid Transformer decoder. We first describe the baseline autoregressive sequence-to-sequence framework and its joint position-value embedding. Building upon this, we introduce Multi-scale Residual Convolutional Embedding (MRCE) and Query-Dependent Gated Attention (QDGA) to tackle the severe misalignments caused by InDel channels. Finally, we analyze the computational complexity of the overall model.

\begin{figure}[t]
    \centering
    \includegraphics[width=\columnwidth]{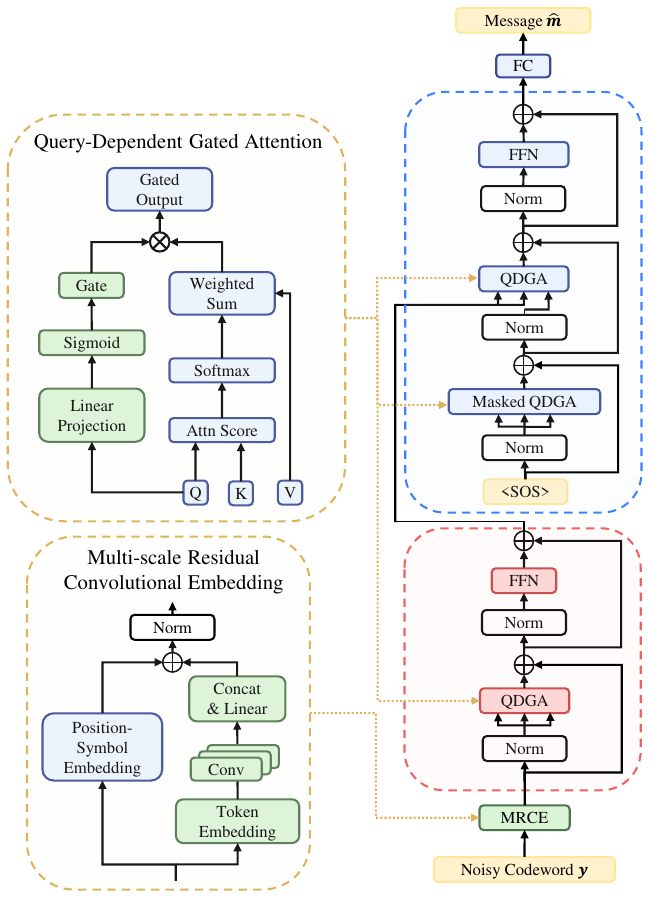}
    \caption{Architecture of the proposed Locality-Aware Hybrid Transformer decoder. MRCE combines position--symbol embeddings with multi-scale convolutional features, while query-dependent gated attention is applied to the encoder self-attention, masked decoder self-attention, and encoder--decoder cross-attention. The noisy codeword is decoded directly into the recovered message.}
    \label{fig:neural_decoder_architecture}
\end{figure}

Figure \ref{fig:neural_decoder_architecture} provides an overview of the proposed neural decoder. The noisy codeword is first transformed by MRCE into a locality-aware representation and then processed by the Transformer encoder and autoregressive decoder, where query-dependent gating modulates the self-attention and cross-attention outputs before the final message prediction.

The foundation of our neural decoder is an autoregressive Transformer. Because InDel errors randomly shift and alter the length of transmitted codewords, the encoder processes received sequences up to a maximum length $n_{\max}$, padding shorter sequences uniformly to enable parallel batch training. A padding mask is subsequently applied in the self-attention layers to prevent interference.

Traditional sinusoidal positional encodings are sub-optimal for heavily misaligned binary sequences. Inspired by \cite{choukroun2022error}, we construct a joint position-symbol statistic embedding. Specifically, we instantiate independent embedding matrices $\mathbf{E}_v \in \mathbb{R}^{n_{\max} \times d_{\text{model}}}$ for each vocabulary token $v \in \{0, 1, \text{pad}\}$. For a received sequence $y = (y_1, \dots, y_m)$, the statistic embedding vector $\mathbf{E}[i, :]$ at the $i$-th position is formulated directly from the corresponding token matrix:
\begin{equation}
\mathbf{E}[i, :] = 
\begin{cases} 
\mathbf{E}_0[i, :] & \text{if } y_i = 0 \\
\mathbf{E}_1[i, :] & \text{if } y_i = 1 \\
\mathbf{E}_{\text{pad}}[i, :] & \text{if } y_i = \text{pad}
\end{cases}.
\end{equation}
This basic embedding $\mathbf{E}$ provides a global positional anchor, upon which our proposed InDel-specific mechanisms are built.

\subsection{Multi-scale Residual Convolutional Embedding}
We propose a Multi-scale Residual Convolutional Embedding (MRCE) to tackle the context-agnostic ambiguity caused by synchronization errors. Unlike traditional scalar embeddings, MRCE extracts local $n$-gram features to provide synchronization-aware sequence representations. 

For a received sequence $\mathbf{y}$, let $\mathbf{E} \in \mathbb{R}^{m \times d_{\text{model}}}$ denote the standard symbol-statistic embedding. The MRCE branch first maps $\mathbf{y}$ into a low-dimensional space $\mathbf{C} \in \mathbb{R}^{m \times d_{\text{cnn}}}$. To capture multi-scale structural shifts, we apply parallel 1D convolutions with varied kernel sizes $\mathcal{S} = \{3, 5\}$:
\begin{equation}
    \mathbf{F}^{(s)} = \text{Conv1D}_s(\mathbf{C}), \quad s \in \mathcal{S}.
\end{equation}
These local feature maps are concatenated and linearly projected back to the original dimension: $\mathbf{H}_{\text{cnn}} = \text{Concat}(\mathbf{F}^{(3)}, \mathbf{F}^{(5)}) \mathbf{W}_{\text{proj}}$. 

Finally, a residual connection fuses the global statistic embedding with the local convolutional features:
\begin{equation}
    \mathbf{E}_{\text{hybrid}} = \text{LayerNorm}(\mathbf{E} + \mathbf{H}_{\text{cnn}}).
\end{equation}
This design provides an explicit anchor for the combinatorial sketches through $\mathbf{E}$, while $\mathbf{H}_{\text{cnn}}$ adaptively learns the spatial perturbations caused by InDel errors. A more detailed discussion of the selection of convolutional kernel parameters is provided in the appendix.

\subsection{Query-Dependent Gated Attention Mechanism}
While MRCE provides robust local features, standard Scaled Dot-Product Attention (SDPA) remains vulnerable to spurious global alignments caused by severe InDel shifts. Inspired by recent studies \cite{qiu2025gated}, we introduce a Query-Dependent Gated Attention to modulate the SDPA output.

For a given attention head with query $Q$, key $K$, and value $V$, the standard output $O = \text{softmax}(QK^T/\sqrt{d_k})V$ is a linear combination of values. We introduce a head-specific gate $G$ projected directly from the query:
\begin{equation}
    G = \sigma(Q W_g), \quad \tilde{O} = O \odot G,
\end{equation}
where $W_g \in \mathbb{R}^{d_k \times d_k}$ is a learnable weight, $\sigma(\cdot)$ is the sigmoid function, and $\odot$ is the element-wise multiplication.

This gating mechanism introduces input-dependent sparsity and non-linearity directly upon the low-rank mapping of SDPA. By modulating the attention outputs, it actively filters out irrelevant noise from misaligned tokens, stabilizing training dynamics and enhancing the decoder's capacity to resolve complex multi-error collisions with negligible parameter overhead.

\subsection{Complexity Analysis}

To demonstrate the computational efficiency of our proposed framework, we analyze the overall time complexity of the hybrid neural decoder. Let $n$ denote the sequence length, $d$ denote the embedding dimension, and $N$ denote the number of transformer layers. 

The baseline autoregressive Transformer is dominated by the Multi-Head Attention and Feed-Forward Networks, yielding a standard complexity of $\mathcal{O}(N(n^2 d + n d^2))$. Our proposed structural innovations, the MRCE module and the Query-Dependent Gated Attention, operate exclusively via 1D convolutions over small kernel sizes and linear gate projections. These operations strictly introduce a marginal linear-time overhead of $\mathcal{O}(nd)$ and $\mathcal{O}(Nndd_k)$ (where head dimension $d_k \ll d$), respectively.

Consequently, the total time complexity of the augmented architecture remains tightly bounded by the baseline:
\begin{equation}
\mathcal{O}\Big(N(n^2 d + n d^2)\Big).
\end{equation}
Importantly, this decoding complexity is entirely independent of the number of errors, standing in stark contrast to traditional algebraic optimal decoders for InDel channels, which suffer from a prohibitive exponential complexity of $\mathcal{O}(n^{t+1})$ for $t$ errors \cite{sima2020optimal}. Thus, our framework efficiently resolves complex multi-error collisions while preserving high scalability for practical DNA storage.

\section{Extension to DNA Storage}

We extend the binary framework to a quaternary IDS channel. Transmitted strands are constrained to have 45\%--55\% GC content and a maximum homopolymer length of five.

\subsection{Quaternary Mapping and Constraints}

For an even binary codeword length $n$, consecutive bit pairs are mapped bijectively to DNA bases:
\begin{equation}
    00 \to A, \quad 01 \to G, \quad 10 \to C, \quad 11 \to T.
\end{equation}
Equivalently, $b_i=2x_{2i}+x_{2i+1}$, so an $n$-bit codeword produces a strand of $L=n/2$ bases. Candidate binary codewords are generated under the selected sketch constraints, mapped to the quaternary alphabet, and retained only when both biochemical constraints are satisfied. The bijective mapping preserves the underlying bit-level constraints.

\subsection{Decoder Adaptation}

The encoder receives only the noisy quaternary sequence, represented by symbols $\{0,1,2,3\}$ with token $4$ reserved for padding, and jointly embeds each symbol and its position in the received sequence. The remaining encoder--decoder architecture is unchanged. The binary output head is replaced by a four-class projection trained with cross-entropy to autoregressively recover the information-base sequence.

For a target information-base sequence $\mathbf{u}=(u_1,\ldots,u_\ell)$ and its prediction $\hat{\mathbf{u}}$, we report the Nucleobase Error Rate (NER):
\begin{equation}
    \mathrm{NER}(\mathbf{u},\hat{\mathbf{u}})
    =\frac{1}{\ell}\sum_{i=1}^{\ell}\mathbf{1}[u_i\neq\hat{u}_i].
\end{equation}

\section{Implementation}
In this section, we evaluate the proposed neural decoder by detailing the experimental setup, presenting the overall error-correction performance, and conducting an architectural ablation study.
\subsection{Experimental Setup}
We evaluate the decoder on independent training and test sets for binary codewords of lengths $n=64$ and $n=128$. Each configuration contains 240,000 training samples and 60,000 test samples. Noisy samples are produced by the InDel or IDS channel specified in each experiment. Channel outputs are padded to the maximum received length, with padding positions masked in attention.

Unless otherwise stated, the Transformer configuration is $d=512$, $d_{\mathrm{FFN}}=2048$, $N_{\mathrm{enc}}=N_{\mathrm{dec}}=3$, and $h=8$ attention heads. MRCE uses a 32-dimensional input representation and convolution kernels $\{3,5\}$. Models are trained from scratch for 1000 epochs using Adam~\cite{kingma2014adam}, an initial learning rate of $10^{-4}$ with cosine annealing to $10^{-6}$, and gradient clipping at a maximum norm of 1.0. During training, the decoder uses teacher forcing; at test time, it recovers the message autoregressively. The random seed is fixed to 0 for all experiments.

\begin{table}[htbp]
    \centering
    \small
    \setlength{\tabcolsep}{1.5mm}
    \renewcommand{\arraystretch}{1.2} 
    \begin{tabular}{ccccc}
        \toprule
        \multirow{2}{*}{\textbf{\shortstack{Total\\Errors}}} & \multicolumn{2}{c}{\textbf{2 Sketches ($f_1, f_2$)}} & \multicolumn{2}{c}{\textbf{3 Sketches ($f_1, f_2, f_3$)}} \\
        \cmidrule(lr){2-3} \cmidrule(lr){4-5}
        & $n=64$ & $n=128$ & $n=64$ & $n=128$ \\
        \midrule
        \multirow{2}{*}{\textbf{0}} 
        & 0.00E-00 & 2.60e-07 & 5.48E-05 & 3.19E-05 \\
        & 0.00E-00 & 1.67e-05 & 8.00E-04 & 0.10\% \\
        \midrule
        \multirow{2}{*}{\textbf{1}} 
        & 7.08E-06 & 2.60e-06 & 3.46E-04 & 0.11\% \\
        & 1.17E-04 & 1.00e-04 & 0.41\% & 2.79\% \\
        \midrule
        \multirow{2}{*}{\textbf{2}} 
        & 0.50\% & 0.12\% & 0.22\% & 0.35\%  \\
        & 2.89\% & 2.22\% & 2.31\% & 7.67\% \\
        \midrule
        \multirow{2}{*}{\textbf{3}} 
        & 9.35\% & 0.74\% & 1.00\% &  0.82\% \\
        & 45.90\% & 11.9\% & 9.46\% & 14.50\% \\
        \midrule
        \multirow{2}{*}{\textbf{4}} 
        & 18.80\% & 2.52\% & 5.03\% & 2.21\% \\
        & 81.20\% & 31.3\% & 35.80\% & 31.40\%  \\
        \bottomrule
    \end{tabular}
    \caption{Bit Error Rate (BER) and Frame Error Rate (FER) under varying numbers of random InDel errors. For each entry, the top value represents the BER and the bottom value represents the FER.}
    \label{tab:main_results}
\end{table}

Training and inference use an NVIDIA RTX 5090 (32 GB) and an NVIDIA RTX 3090 Ti (24 GB). The host hardware includes an AMD EPYC 9754 128-Core Processor, an Intel Xeon Gold 5220R CPU, and 503 GiB of system memory. The software environment is Ubuntu 22.04.5 LTS with Python 3.13.2 as the primary programming language.

\subsection{Error-Correction Performance}

Table \ref{tab:main_results} presents the decoding performance of our best-performing models on the binary InDel channel, measured in both Bit Error Rate (BER) and Frame Error Rate (FER), across codeword lengths $n=64$ and $n=128$ with varying numbers of combinatorial sketches.

As shown in Table \ref{tab:main_results}, a clear scalable correspondence emerges between the number of sketches and the error-correction capability. With two sketches ($f_1, f_2$), the decoder suppresses BER below 1\% under two concurrent InDel errors (0.50\% for $n=64$, 0.12\% for $n=128$). Augmenting the codeword with the third-order sketch ($f_3$) extends this capability by one additional error: under three concurrent errors, BER remains below 1\% (0.998\% for $n=64$, 0.82\% for $n=128$). This property holds consistently across both codeword lengths, validating the scalability of combining higher-order sketches with neural decoding.
\begin{table}[htbp]
    \centering
    \small
    \setlength{\tabcolsep}{1.5mm}
    \renewcommand{\arraystretch}{1.2}
    \begin{tabular}{ccccc}
        \toprule
        \multirow{2}{*}{\textbf{\shortstack{Total\\Errors}}} & \multicolumn{2}{c}{\textbf{2 Sketches ($f_1, f_2$)}} & \multicolumn{2}{c}{\textbf{3 Sketches ($f_1, f_2, f_3$)}} \\
        \cmidrule(lr){2-3} \cmidrule(lr){4-5}
        & $n=64$ & $n=128$ & $n=64$ & $n=128$ \\
        \midrule
        \multirow{2}{*}{\textbf{0}} 
        & 0.00E+00 & 6.61E-04 & 2.08E-05 & 1.16E-06 \\
        & 0.00E+00 & 1.90\% & 2.67E-04 & 5.00E-05 \\
        \midrule
        \multirow{2}{*}{\textbf{1}} 
        & 1.29E-05 & 0.38\% & 9.29E-04 & 2.79E-04 \\
        & 3.67E-04 & 10.40\% & 1.45\% & 0.46\% \\
        \midrule
        \multirow{2}{*}{\textbf{2}} 
        & 1.74\% & 1.19\% & 0.77\% & 0.27\% \\
        & 9.47\% & 25.10\% & 9.08\% & 3.47\% \\
        \midrule
        \multirow{2}{*}{\textbf{3}} 
        & 13.60\% & 2.71\% & 2.94\% & 1.07\% \\
        & 71.50\% & 43.30\% & 28.00\% & 13.50\% \\
        \midrule
        \multirow{2}{*}{\textbf{4}} 
        & 19.80\% & 5.07\% & 8.54\% & 2.79\% \\
        & 91.70\% & 61.70\% & 60.90\% & 33.20\% \\
        \bottomrule
    \end{tabular}
    \caption{Bit Error Rate (BER) and Frame Error Rate (FER) under varying numbers of random errors in the IDS (Insertion, Deletion, Substitution) channel. For each entry, the top value represents the BER and the bottom value represents the FER.}
    \label{tab:ids_results}
\end{table}

We further evaluated the framework on the IDS channel, which additionally introduces substitution errors. As shown in Table \ref{tab:ids_results}, the third-order sketch provides its largest gains as the number of channel errors increases. Under three errors, adding $f_3$ reduces the BER from 13.60\% to 2.94\% for $n=64$, and from 2.71\% to 1.07\% for $n=128$. 

For an external reference point, Table~\ref{tab:vtformer_comparison} compares our results with the reported VT-Former results~\cite{wei2025vtformer} at matched codeword lengths $n=68$ and $n=100$. The reported accuracy values are converted to error rates to provide a consistent comparison in which lower values are better.

\begin{table}[htbp]
    \centering
    \small
    \setlength{\tabcolsep}{1.0mm}
    \renewcommand{\arraystretch}{1.2}
    \begin{tabular}{ccccc}
        \toprule
        \multirow{2}{*}{\textbf{\shortstack{Total\\Errors}}} & \multicolumn{2}{c}{$n=68$} & \multicolumn{2}{c}{$n=100$} \\
        \cmidrule(lr){2-3} \cmidrule(lr){4-5}
        & \textbf{VT-Former} & \textbf{Ours} & \textbf{VT-Former} & \textbf{Ours} \\
        \midrule
        \multirow{2}{*}{\textbf{1}}
        & 0.00\% & $<0.01\%$ & 0.00\% & 0.01\% \\
        & 0.00\% & 0.02\% & 0.00\% & 0.13\% \\
        \midrule
        \multirow{2}{*}{\textbf{2}}
        & 16.30\% & 0.08\% & 15.90\% & 0.21\% \\
        & 86.20\% & 0.54\% & 91.80\% & 2.23\% \\
        \midrule
        \multirow{2}{*}{\textbf{3}}
        & 21.90\% & 1.85\% & 21.80\% & 3.28\% \\
        & 96.80\% & 10.00\% & 97.60\% & 22.70\% \\
        \midrule
        \multirow{2}{*}{\textbf{4}}
        & 25.00\% & 9.45\% & 25.50\% & 11.30\% \\
        & 99.40\% & 48.20\% & 99.80\% & 67.50\% \\
        \bottomrule
    \end{tabular}
    \caption{BER and FER comparison with the reported VT-Former results~\cite{wei2025vtformer} on the binary IDS channel. For each entry, the top value represents BER and the bottom value represents FER.}
    \label{tab:vtformer_comparison}
\end{table}

Our decoder achieves markedly lower BER under multiple errors. With three errors, BER decreases from 21.90\% to 1.85\% at $n=68$ and from 21.80\% to 3.28\% at $n=100$. Even with four errors, our BER remains at 9.45\% and 11.30\%, compared with 25.00\% and 25.50\% for VT-Former, respectively. Because the results were obtained using the respective code constructions and evaluation pipelines, they should be interpreted as a reference comparison at matched codeword lengths and total error counts rather than as an identical-protocol benchmark.

\subsection{Extension to DNA-Constrained Quaternary Channel}

To validate that our framework generalizes beyond the binary domain, we tested the quaternary model described in Section 6 on the IDS channel with GC balance and run-length constraints enforced. We use binary codewords of lengths $n=128$ and $n=300$ (corresponding to 64 and 150 bases, respectively) with the three-sketch configuration ($f_1, f_2, f_3$). Table \ref{tab:dna_results} reports the Nucleobase Error Rate (NER) and Frame Error Rate (FER).

\begin{table*}[!t]
    \centering
    \renewcommand{\arraystretch}{1.0}
    \small
    \setlength{\tabcolsep}{1.2mm}
    \begin{tabular}{cccccccccccc}
        \toprule
        \multicolumn{2}{c}{\textbf{Modules}} & \multicolumn{5}{c}{\textbf{Codeword Length $n=64$}} & \multicolumn{5}{c}{\textbf{Codeword Length $n=128$}} \\
        \cmidrule(lr){1-2} \cmidrule(lr){3-7} \cmidrule(lr){8-12}
        \textbf{MRCE} & \textbf{QDGA} & \textbf{0 Error} & \textbf{1 Error} & \textbf{2 Errors} & \textbf{3 Errors} & \textbf{4 Errors} & \textbf{0 Error} & \textbf{1 Error} & \textbf{2 Errors} & \textbf{3 Errors} & \textbf{4 Errors} \\
        \midrule
        & & 4.88E-05 & 0.28\%  & 2.37\%  & 7.12\%  & 13.00\% & 0.00E-00 & 9.49E-06 & 4.47E-04 & 0.44\% & 1.79\% \\
        & & 5.83E-04 & 3.54\% & 33.10\% & 69.60\% & 89.10\% & 0.00E-00 & \textbf{2.67E-04} & 0.71\% & 6.60\% & 25.20\% \\
        \midrule
        \checkmark & & \textbf{1.90E-05} & \textbf{2.32E-05} & \textbf{4.24E-04} & \textbf{0.49\%}  & \textbf{3.94\%} & 7.31E-05 & 0.15\%  & 0.47\% & 0.99\% & 2.47\%  \\
        & & \textbf{1.67E-04} & \textbf{2.00E-04} & \textbf{0.49\%}  & \textbf{4.44\%}  & \textbf{29.30\%} & 0.30\%  & 4.32\% & 11.70\% & 20\%  & 39.50\% \\
        \midrule
        & \checkmark & 5.65E-05 & 0.29\% & 2.40\% & 7.29\% & 13.10\% & \textbf{0.00E-00} & \textbf{7.41E-06} & \textbf{3.41E-04} & \textbf{0.40\%}& \textbf{1.66\%} \\
        & & 4.17E-04 & 3.92\% & 34.10\% & 71.80\% & 89.70\% & \textbf{0.00E-00} & 3.33E-04 & \textbf{0.56\%} & \textbf{5.70\%} & \textbf{22.30\%} \\
        \midrule
        \checkmark & \checkmark & 5.48E-05 & 3.46E-04 & 0.22\% & 1.00\%  & 5.03\% &   3.19E-05 & 0.10\% & 0.35\%  & 0.82\% & 2.21\%  \\
        & & 8.00E-04 & 0.41\% & 2.31\%  & 9.46\%  & 35.80\% & 0.10\% & 2.79\% & 7.67\%  & 14.50\% & 31.40\% \\
        \bottomrule
    \end{tabular}
        \caption{Ablation study on the MRCE and QDGA using the 3-sketch configuration. For each entry, the top value represents the BER and the bottom value represents the FER. Best results in each column are highlighted in bold.}
    \label{tab:ablation}
\end{table*}

\begin{table}[htbp]
    \centering
    \small
    \setlength{\tabcolsep}{1.2mm}
    \renewcommand{\arraystretch}{1.2}
    \begin{tabular}{ccccc}
        \toprule
        \multirow{2}{*}{\textbf{\shortstack{Total\\Errors}}} & \multicolumn{2}{c}{$n=128$ (64 bases)} & \multicolumn{2}{c}{$n=300$ (150 bases)} \\
        \cmidrule(lr){2-3} \cmidrule(lr){4-5}
        & \textbf{NER} & \textbf{FER} & \textbf{NER} & \textbf{FER} \\
        \midrule
        \textbf{0} & 6.16E-04 & 1.19\% & 1.89E-04 & 0.88\% \\
        \textbf{1} & 0.57\% & 9.84\% & 0.82\% & 33.40\% \\
        \textbf{2} & 1.78\% & 23.90\% & 2.18\% & 63.80\% \\
        \textbf{3} & 4.84\% & 49.90\% & 4.53\% & 87.20\% \\
        \textbf{4} & 9.33\% & 74.60\% & 7.39\% & 96.90\% \\
        \bottomrule
    \end{tabular}
    \caption{NER and FER on the DNA-constrained quaternary IDS channel with three sketches. All strands satisfy 45\%--55\% GC content and a maximum run length of 5.}
    \label{tab:dna_results}
\end{table}

As shown in Table \ref{tab:dna_results}, our framework maintains meaningful error-correction capability under the full set of biological constraints. For the longer $n=300$ codewords (150 bases), the NER increases from 0.82\% under one error to 7.39\% under four errors. The performance degradation compared to the binary IDS results (Table \ref{tab:ids_results}) is expected, as each decoding step now faces a 4-class classification problem and the halved sequence length reduces the structural redundancy available to the decoder. Nevertheless, these results confirm that the proposed sketch-based encoding and neural decoding architecture can be successfully extended to operate directly on the quaternary DNA alphabet with practical biochemical constraints.

\subsection{Ablation Study}

To isolate the contributions of MRCE and QDGA, we evaluate the baseline, each module individually, and their combination under the three-sketch setting for $n=64$ and $n=128$. Table \ref{tab:ablation} reports BER and FER across zero to four InDel errors.

As shown in Table \ref{tab:ablation}, the baseline Transformer architecture (without MRCE and without QDGA) generally yields the highest error rates across almost all configurations, validating the necessity of specialized structural designs for InDel channels. Integrating either the MRCE module or the QDGA mechanism independently results in significant performance gains. Notably, the MRCE module provides substantial improvements in mitigating local synchronization drifts, as evidenced by the sharp drop in BER (e.g., from 7.12\% to 0.49\% for $n=64$ with 3 errors).

While the integration of both modules (MRCE + QDGA) yields highly competitive results, a granular inspection reveals that it does not strictly dominate the performance matrix across every single scenario. For instance, under the severe condition of 4 errors at $n=64$, the MRCE-only model achieves a BER of 3.94\% and an FER of 29.30\%, outperforming the combined model. Similarly, at $n=128$ with 3 and 4 errors, the Gate-only configuration surprisingly delivers the most optimal BERs (0.40\% and 1.66\%, respectively).

Despite these numerical fluctuations under different codeword length, we ultimately selected the combined structure (MRCE + QDGA) as our final deployed architecture. This design choice is rooted in overarching system robustness. By coupling them, the model achieves favorable performance across different scenario configurations.

\section{Conclusion}
In this paper, we introduced a neural decoding framework for multiple insertion and deletion (InDel) errors that combines higher-order combinatorial sketches with a Transformer-based decoder. The encoder operates directly on noisy code sequences, and the autoregressive decoder recovers the corresponding message. MRCE aggregates local context at multiple scales, while QDGA modulates each attention output according to its query. These modules do not change the Transformer's leading-order decoding complexity.

On binary InDel and IDS channels, additional sketches generally improve decoding as the number of errors increases. In the evaluated configurations, each added sketch extends by one error the range in which BER remains below 1\% on InDel channel. At matched lengths and error counts, our models report lower BER than VT-Former for two to four errors, although the methods use different code constructions and evaluation pipelines. The ablation results also show that MRCE and QDGA each contribute to performance.

We also adapted the decoder to DNA sequences subject to GC-content and homopolymer constraints. Results for sequences of 64 and 150 bases show that the same decoding pipeline can be applied to the constrained alphabet, although error rates increase as more IDS operations are introduced. Future work will examine longer sequences, improve constraint-aware training, and reduce the performance gap between binary and quaternary decoding.

\section{Additional Details and Experiments}

Four supplementary analyses are provided in Appendices~\ref{sec:encoding_rate}--\ref{sec:error_types}. We first describe the codeword-search and dataset-construction procedure and estimate the empirical encoding rate through Monte Carlo sampling. We then examine the effects of sketch moduli and the MRCE convolutional branch design. Finally, we evaluate the decoder separately under insertion, deletion, and substitution errors to assess its sensitivity to error type. Together, these analyses clarify the encoding procedure, effective rate, sketch design, MRCE configuration, and behavior under individual channel disturbances.

\clearpage
\appendix
\section*{Appendices}
\section{Codeword Search, Dataset Construction, and Empirical Encoding Rate}
\label{sec:encoding_rate}

This section motivates the evaluated codeword lengths, provides implementation details for codeword construction, and quantifies its operational encoding rate. Section~\ref{app:method1} explains the choice of the binary block lengths and DNA strand lengths. Section~\ref{app:method2} defines the sketch functions, moduli, and notation for the information and search positions in the binary and DNA configurations. Section~\ref{app:method3} describes the meet-in-the-middle precomputation and budgeted depth-first search used to complete a message into a valid codeword, including the DNA constraints. Finally, Section~\ref{app:method4} defines the operational encoding rate, explains its Monte Carlo estimation, and reports the estimates for all evaluated datasets.

\subsection{Rationale for the Evaluated Lengths}\label{app:method1}

Our primary binary evaluations use $n\in\{64,128\}$, two consecutive powers of two. In the three-sketch setting, $M_j=n^j$, so dyadic $n$ makes $\log_2 M_j=j\log_2 n$ integral, following a related deletion-code analysis that adopts the same convention to avoid ceiling and floor terms~\cite{gabrys2022beyond}. This simplifies nominal redundancy accounting but does not restrict the construction: the $n=68$ and $n=100$ experiments are included to match the settings reported by VT-Former~\cite{wei2025vtformer}. For DNA, $n$ denotes the binary representation length and $L=n/2$ the strand length, so $n=300$ corresponds to $L=150$ nucleotides. Since prior work reports Illumina sequence lengths of 100--300 DNA symbols~\cite{maarouf2022concatenated}, 150 nucleotides provides a representative practical setting alongside the shorter 64-nucleotide configuration rather than a theoretically optimal length.

\subsection{Sketch Definitions and Codeword Configurations}\label{app:method2}

This section details the codeword-search procedure introduced in Section~4.2 of the main paper. We use zero-based coordinates $p\in\{0,\ldots,n-1\}$ to match the implementation. For the three-sketch setting, define
\begin{equation}
\begin{aligned}
    a_j(p)&=\binom{p}{j},\\
    f_j(\mathbf{x})&=\sum_{p=0}^{n-1}a_j(p)x_p\pmod{M_j},
    \qquad j\in\{1,2,3\}.
\end{aligned}
\end{equation}
The generators use $(M_1,M_2,M_3)=(n,n^2,n^3)$ and the target residues $f_j(\mathbf{x})=0$. Let $\mathcal{I}$ be the fixed set of $k$ information positions and let $\mathcal{P}=\mathcal{I}^{c}$ contain the $n-k$ positions determined by the search. We partition $\mathcal{P}$ into disjoint ordered sets $\mathcal{P}_{L}$ and $\mathcal{P}_{R}$.

The two-sketch configurations use the same partitioned search after omitting $f_3$, $M_3$, and the third residue coordinate from the hash key and lookup. Their moduli are specified in Section~\ref{sec:modulus_ablation}.

For DNA codewords, the two bits representing each nucleotide are assigned together to either the information set or the search set.

\subsection{Meet-in-the-Middle Codeword Search}\label{app:method3}

For an assignment $\mathbf{l}$ on $\mathcal{P}_{L}$, let
\begin{equation}
    \rho_j(\mathbf{l})=
    \sum_{p\in\mathcal{P}_{L}}a_j(p)l_p\pmod{M_j},
\end{equation}
and define $\rho_j(\mathbf{r})$ analogously on $\mathcal{P}_{R}$. The encoder first enumerates the $2^{|\mathcal{P}_{L}|}$ left assignments once and stores them in a hash table indexed by the residue triple $(\rho_1,\rho_2,\rho_3)$. This table is reused for every message. Algorithm~\ref{alg:dataset_generation} gives the common search and data-generation procedure.

\begin{algorithm*}[t]
\caption{Three-sketch codeword search and dataset construction}
\label{alg:dataset_generation}
\textbf{Input}: $\mathcal{I}$, $\mathcal{P}_{L}$, $\mathcal{P}_{R}$, moduli $M_1,M_2,M_3$, target size $N$, mode $q\in\{\mathrm{binary},\mathrm{DNA}\}$, and per-message search budget $B_q$\\
\textbf{Output}: A set $\mathcal{D}$ of valid codewords
\begin{algorithmic}[1]
\STATE Initialize an empty hash table $H$.
\FOR{each $\mathbf{l}\in\{0,1\}^{|\mathcal{P}_{L}|}$}
    \STATE Append $\mathbf{l}$ to $H[(\rho_1(\mathbf{l}),\rho_2(\mathbf{l}),\rho_3(\mathbf{l}))]$.
\ENDFOR
\STATE $\mathcal{D}\gets\varnothing$ and $\mathcal{U}\gets\varnothing$.
\WHILE{$|\mathcal{D}|<N$ and the sampling budget remains}
    \STATE Draw a message $\mathbf{m}$ using the rule for mode $q$.
    \STATE Set $\mathbf{x}_{\mathcal{I}}\gets\mathbf{m}$ and all search positions to zero.
    \STATE $T_j\gets\sum_{p\in\mathcal{I}}a_j(p)x_p\pmod{M_j}$ for $j=1,2,3$.
    \STATE $\widehat{\mathbf{x}}\gets\bot$.
    \FOR{each right leaf $\mathbf{r}$ reached by zero-first DFS, counting every visited node toward $B_q$}
        \STATE $d_j\gets-(T_j+\rho_j(\mathbf{r}))\pmod{M_j}$ for $j=1,2,3$.
        \FOR{each $\mathbf{l}\in H[(d_1,d_2,d_3)]$}
            \STATE Assemble $\mathbf{x}$ from $\mathbf{m}$, $\mathbf{l}$, and $\mathbf{r}$.
            \IF{$q=\mathrm{binary}$ or $\mathbf{x}$ satisfies the DNA constraints}
                \STATE $\widehat{\mathbf{x}}\gets\mathbf{x}$; \textbf{break both loops}.
            \ENDIF
        \ENDFOR
    \ENDFOR
    \IF{$\widehat{\mathbf{x}}\neq\bot$ and $\mathbf{m}\notin\mathcal{U}$}
        \STATE $\mathcal{D}\gets\mathcal{D}\cup\{\widehat{\mathbf{x}}\}$ and $\mathcal{U}\gets\mathcal{U}\cup\{\mathbf{m}\}$.
    \ENDIF
\ENDWHILE
\STATE \textbf{return} $\mathcal{D}$
\end{algorithmic}
\end{algorithm*}

Given a message, both generators traverse $\mathcal{P}_{R}$ using zero-first depth-first search and stop after a prescribed number of visited recursion nodes. The nominal per-message budgets are $B_{\mathrm{binary}}=5\times10^5$ and $B_{\mathrm{DNA}}=2\times10^6$. At each complete right assignment, the required left residue triple is queried from the hash table. If its bucket contains several assignments, the binary mode uses the first one, whereas the DNA mode checks them in stored order until the complete strand also satisfies the biological constraints. Both modes return the first valid completion found. Consequently, a search failure means that no valid codeword was found within the allocated budget, rather than proving that no valid assignment exists.

In DNA mode, adjacent bit pairs are mapped as $00\mapsto A$, $01\mapsto G$, $10\mapsto C$, and $11\mapsto T$. A candidate is accepted only if G/C bases constitute 45\%--55\% of the complete strand and no homopolymer is longer than five bases. As a search-efficiency heuristic, the information-message sampler applies a GC-content prefilter to exclude highly unbalanced proposals. The complete-strand constraints are still checked after a sketch match.

For dataset construction, binary messages are sampled uniformly from $\{0,1\}^{k}$. Only a previously unseen message for which the search succeeds is retained, and one full $n$-bit codeword is written per line. The current generators collect a pool of $N=300{,}000$ codewords; the disjoint training and test sets described in the main paper are formed from this pool outside the C++ search routine. Multiple threads sample and search independently, while insertion into the accepted-message set is synchronized to prevent duplicates.

\subsection{Monte Carlo Estimate of the Operational Encoding Rate}\label{app:method4}

Let $\mathcal{A}\subseteq\{0,1\}^{k}$ denote the messages for which the encoder returns a valid codeword, including the biological constraints when applicable. If $p_{\mathrm{enc}}=|\mathcal{A}|/2^k$, then the effective binary rate is
\begin{equation}
    R_{\mathrm{eff}}=\frac{\log_2|\mathcal{A}|}{n}
    =\frac{k+\log_2 p_{\mathrm{enc}}}{n},
\end{equation}
which is below the nominal rate $k/n$ whenever $p_{\mathrm{enc}}<1$. In practice, each dataset generator draws random message proposals until $S=300{,}000$ distinct messages have been encoded successfully. If $T$ denotes the recorded number of proposals, we compute
\begin{equation}
    \widehat{p}_{\mathrm{op}}=\frac{S}{T},\qquad
    \widehat{R}_{\mathrm{op}}=
    \frac{k+\log_2\widehat{p}_{\mathrm{op}}}{n}.
\end{equation}
For binary datasets, proposals are sampled uniformly from $\{0,1\}^k$. The DNA sampler first applies its information-GC proposal filter; hence, its $\widehat{p}_{\mathrm{op}}$ is conditional on that sampling rule. Moreover, all generators use a finite right-search budget. We therefore report $\widehat{R}_{\mathrm{op}}$ as an operational rate estimate of the implemented dataset-construction process, rather than an exact codebook rate over the full message space.

Table~\ref{tab:encoding_rate} summarizes the results for every dataset configuration used in the main paper. We include the nominal rate $R_{\mathrm{nom}}=k/n$ for comparison.

\begin{table}[htbp]
    \centering
    \footnotesize
    \renewcommand{\arraystretch}{1.12}
    \setlength{\tabcolsep}{0.65mm}
    \begin{tabular}{ccrrrr}
        \toprule
        $s$ & $(n,k)$ & $T$ & $p_{\mathrm{op}}$ & $R_{\mathrm{nom}}$ & $R_{\mathrm{op}}$ \\
        \midrule
        \multicolumn{6}{l}{\textbf{Binary}} \\
        2 & $(64,40)$  & $1{,}114{,}580$   & 26.92\%  & 0.6250 & 0.5954 \\
        2 & $(128,64)$ & $5{,}794{,}369$   & 5.18\%   & 0.5000 & 0.4666 \\
        3 & $(64,36)$  & $5{,}784{,}724$   & 5.19\%   & 0.5625 & 0.4958 \\
        3 & $(68,40)$  & $904{,}317{,}943$ & 0.0332\% & 0.5882 & 0.4183 \\
        3 & $(100,52)$ & $750{,}754$       & 39.96\%  & 0.5200 & 0.5068 \\
        3 & $(128,72)$ & $3{,}375{,}112$   & 8.89\%   & 0.5625 & 0.5352 \\
        \midrule
        \multicolumn{6}{l}{\textbf{DNA}} \\
        3 & $(128,64)$  & $10{,}786{,}396$ & 2.78\% & 0.5000 & 0.4596 \\
        3 & $(300,228)$ & $4{,}385{,}501$  & 6.84\% & 0.7600 & 0.7471 \\
        \bottomrule
    \end{tabular}
    \caption{Empirical operational encoding rates. Here, $s=2$ denotes $(f_1,f_2)$, $s=3$ denotes $(f_1,f_2,f_3)$, and every row has $S=300{,}000$ successful encodings. $T$ is the recorded number of proposals and $\widehat{p}_{\mathrm{op}}=S/T$. For DNA, $n$ is the binary representation length and the acceptance probability is conditional on the information-GC proposal filter.}
    \label{tab:encoding_rate}
\end{table}

The operational rates are lower than their nominal counterparts by 1.29--17.00 percentage points. The largest gap occurs for the length-$68$ three-sketch configuration, whose measured acceptance probability is only $0.0332\%$ under the current search policy and budget. For the DNA configurations, the operational rates 0.4596 and 0.7471 correspond to approximately 0.92 and 1.49 information bits per nucleotide, respectively.

\section{Impact of Sketch Moduli on Decoding Performance}
\label{sec:modulus_ablation}

In algebraic coding theory, the moduli used in syndrome calculations affect the collision rate and analytical decodability of a code. To examine their effect in our learning-based framework, we evaluate the baseline Transformer on length-$128$ codewords using the three-sketch configuration $(f_1,f_2,f_3)$. Four modulus sets $(q_1,q_2,q_3)$, denoted by M1--M4, are compared in Table~\ref{tab:modulus_ablation}.

\begin{table}[htbp]
    \centering
    \renewcommand{\arraystretch}{1.2}
    \small
    \setlength{\tabcolsep}{2mm}
    \begin{tabular}{ccccc}
        \toprule
        \textbf{Errors} & \textbf{M1} & \textbf{M2} & \textbf{M3} & \textbf{M4} \\
        \midrule
        \multirow{2}{*}{\textbf{0}}
        & 0.00E-00 & 0.00E-00 & 0.00E-00 & 0.00E-00 \\
        & 0.00E-00 & 0.00E-00 & 0.00E-00 & 0.00E-00 \\
        \midrule
        \multirow{2}{*}{\textbf{1}}
        & 1.07E-05 & 1.85E-05 & 2.82E-05 & 4.28E-05 \\
        & 1.02E-04 & 2.17E-04 & 5.50E-04 & 8.50E-04 \\
        \midrule
        \multirow{2}{*}{\textbf{2}}
        & 0.32\% & 0.26\% & 0.27\% & 0.32\% \\
        & 4.85\% & 3.96\% & 4.34\% & 5.33\% \\
        \midrule
        \multirow{2}{*}{\textbf{3}}
        & 2.67\% & 2.64\% & 2.40\% & 2.98\% \\
        & 30.70\% & 30.60\% & 28.90\% & 33.00\% \\
        \midrule
        \multirow{2}{*}{\textbf{4}}
        & 11.00\% & 10.31\% & 10.50\% & 11.70\% \\
        & 81.10\% & 78.60\% & 79.10\% & 82.10\% \\
        \bottomrule
    \end{tabular}
    \caption{Decoding performance under different sketch modulus sets for $n=128$ with three sketches using the baseline model. M1: $(n,n^2,n^3)$; M2: $(2n{+}1,n^2,n^3)$; M3: $(n,2n^2{+}2,n^3)$; M4: $(n,n^2{-}2n{+}2,n^3)$. For each entry, the top value is BER and the bottom value is FER.}
    \label{tab:modulus_ablation}
\end{table}

\paragraph{Moduli used in the main experiments.}
Table~\ref{tab:modulus_ablation} shows only modest variation across the evaluated three-sketch configurations. We therefore use
$(M_1,M_2,M_3)=(n,n^2,n^3)$ for all three-sketch experiments.

For the two-sketch setting, we choose the modulus scale according to the span of the sketch perturbations induced by up to two InDel operations. Following the one-based positional convention in the main paper, the two extremes are obtained by deleting two terminal ones and by inserting two ones after the terminal position. The corresponding spans for the first- and second-order sketches are
\begin{align}
    D_1+I_1
    &= [(n-1)+n]+[(n+1)+(n+2)] = 4n+2, \\
    D_2+I_2
    &= \left[\binom{n-1}{2}+\binom{n}{2}\right] \notag\\
    &\quad+\left[\binom{n+1}{2}+\binom{n+2}{2}\right] \notag\\
    &= (n-1)^2+(n+1)^2 = 2n^2+2.
\end{align}
These spans motivate the initial choices $M_1=4n+2$ and $M_2=2n^2+2$. Preliminary experiments with the latter second-order modulus yielded weaker decoding performance, so we applied a factor-of-two safety margin to $M_2$. Accordingly, all two-sketch experiments use
$(M_1,M_2)=(4n+2,4n^2+4)$. This choice is an empirical design rule for the neural decoder rather than a claim of the minimum algebraic moduli.

\FloatBarrier
\section{Ablation of MRCE Convolutional Branches}
\label{sec:mrce_branches}

We ablate the convolutional branches of MRCE on the binary InDel channel using $n=64$, three sketches $(f_1,f_2,f_3)$ with moduli $(n,n^2,n^3)$, and QDGA enabled in every model. The only change is the set of convolutional kernel sizes, as reported in Table~\ref{tab:mrce_kernel_ablation}.

\begin{center}
\begin{minipage}{\columnwidth}
    \centering
    \small
    \renewcommand{\arraystretch}{1.12}
    \setlength{\tabcolsep}{2.0mm}
    \begin{tabular}{cccc}
        \toprule
        \textbf{Errors} & $\{3\}$ & $\{3,5\}$ & $\{3,5,7\}$ \\
        \midrule
        \multirow{2}{*}{\textbf{0}}
        & 1.01E-04 & 5.48E-05 & 1.52E-04 \\
        & 0.15\% & 8.00E-04 & 0.22\% \\
        \midrule
        \multirow{2}{*}{\textbf{1}}
        & 0.16\% & 3.46E-04 & 3.02E-04 \\
        & 1.53\% & 0.41\% & 0.40\% \\
        \midrule
        \multirow{2}{*}{\textbf{2}}
        & 0.43\% & 0.22\% & 0.23\% \\
        & 4.56\% & 2.31\% & 2.63\% \\
        \midrule
        \multirow{2}{*}{\textbf{3}}
        & 1.24\% & 1.00\% & 0.90\% \\
        & 12.00\% & 9.46\% & 8.14\% \\
        \midrule
        \multirow{2}{*}{\textbf{4}}
        & 5.68\% & 5.03\% & 4.61\% \\
        & 40.60\% & 35.80\% & 32.40\% \\
        \bottomrule
    \end{tabular}
    \captionof{table}{MRCE kernel-branch ablation on the binary InDel channel. For each error count, the top and bottom entries are BER and FER, respectively.}
    \label{tab:mrce_kernel_ablation}
\end{minipage}
\end{center}

Using only the kernel-$3$ branch is consistently weaker, while adding the kernel-$5$ branch substantially improves both BER and FER. Adding a kernel-$7$ branch improves both metrics at one, three, and four errors, but the gains are not uniform: both metrics degrade slightly at two errors. More importantly, under the zero-error setting, BER and FER increase by factors of 2.77 and 2.71, respectively. Since an uncorrupted codeword should be recovered without introducing errors, and the third branch also increases the parameter count and computation, we select the two-branch configuration $\{3,5\}$ as the better reliability--complexity trade-off.

\section{Performance under Individual Error Types}
\label{sec:error_types}

We evaluate the same $n=128$ three-sketch checkpoint used for the mixed IDS results in the main paper. The checkpoint is trained on the mixed IDS channel and is tested, without retraining or error-type-specific fine-tuning, under four channel settings: mixed IDS, insertion only, deletion only, and substitution only. In every setting, the simulator executes $t\in\{0,1,2,3,4\}$ channel operations per sample. Thus, $t$ denotes the number of executed operations rather than the realized edit distance between the transmitted and received sequences.

\begin{figure}[t]
    \centering
    \includegraphics[width=\columnwidth]{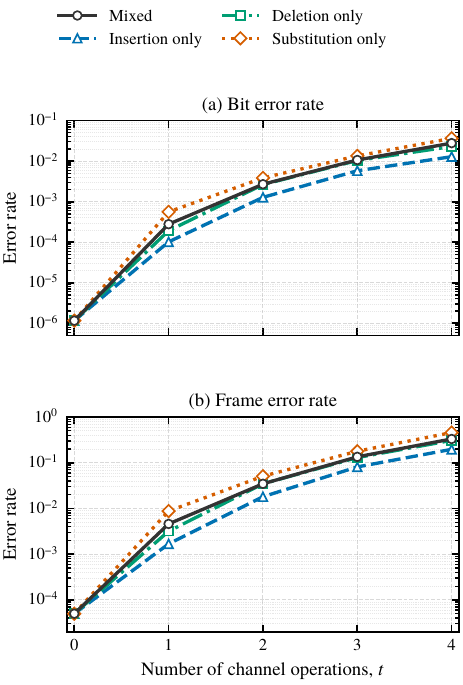}
    \caption{Error-type sensitivity of the same mixed-IDS-trained $n=128$ three-sketch decoder. The model is evaluated without error-type-specific fine-tuning. The upper and lower panels report BER and FER, respectively, and the horizontal axis gives the number of executed channel operations rather than the realized edit distance.}
    \label{fig:error_type_performance}
\end{figure}

As shown in Figure~\ref{fig:error_type_performance}, insertion-only corruption yields the lowest BER and FER for every $t>0$, whereas substitution-only corruption yields the highest. The deletion-only results closely track those of the mixed channel. At $t=4$, for example, the BERs are $1.29\%$, $2.30\%$, $2.79\%$, and $3.66\%$ for the insertion-only, deletion-only, mixed, and substitution-only channels, respectively; the FER curves exhibit the same ordering. All curves coincide at $t=0$ because no corruption is applied. Since insertion, deletion, and substitution operations are all represented in the mixed training channel, this experiment characterizes the error-type sensitivity of a fixed decoder rather than zero-shot generalization to unseen corruption types.

\FloatBarrier

\clearpage
\bibliography{aaai2027}

\end{document}